\documentclass[a4paper]{article}
\usepackage[utf8]{inputenc}
\usepackage{graphicx}
\usepackage{graphicx,wrapfig,lipsum}
\usepackage{array}
\usepackage[colorlinks=true,linkcolor=black, citecolor=blue]{hyperref}%
\usepackage{float, colortbl}
\usepackage{sans}
\usepackage{amsmath}

\usepackage{xcolor}
\usepackage{dirtree}
\usepackage{newpxtext}
\usepackage{newpxmath}
\usepackage{bm}
\usepackage[caption=false]{subfig}
\usepackage[font={footnotesize},labelfont={bf}]{caption}
\usepackage{xcolor}
\newcommand{\beq}{\begin{equation}}
\newcommand{\eeq}{\end{equation}}
\newcommand{\bea}{\begin{eqnarray}}
\newcommand{\eea}{\end{eqnarray}}
\newcommand{\beqa}{\begin{eqnarray}}
\newcommand{\eeqa}{\end{eqnarray}}

\newlength{\dinwidth}
\newlength{\dinmargin}
\title{Beyond Linearity: Full-scale response modelling and experimental validation of LVDTs}
\author{
 $\rm Kumar \ Akhil \ Kukkadapu^{a}$ \\ \\
$^a${\it  University of Antwerp, Particle Physics group,}\\ 
    {\it Groenenborgerlaan 171, 2020 Antwerpen, Belgium} \\ \\
}

\date{}

\begin{document}
\maketitle

%--------------------------------------------------------------------------------------------------------
\begin{abstract}
Linear Variable Differential Transformers (LVDTs) are widely used as high-precision, contactless displacement sensors in industrial, metrological, and scientific applications. Their performance is typically characterised only within the central linear operating region, whereas their behaviour across large displacements, where geometric and electromagnetic effects introduce significant non-linearity, remains poorly researched. In this work, we present a comprehensive analysis of the full-range response of an LVDT, spanning its entire mechanical stroke. Using a combination of custom-developed finite-element modelling pipeline and dedicated laboratory measurements, we demonstrate that the LVDT response comprises several distinct dynamical regimes. We introduce an analytically unified expression that accurately reproduces the measured response over all displacement scales, including linear and non-linear ranges, capturing the underlying physical features of the device. The model achieves high fidelity across the full range and provides closed-form first and second derivatives, enabling unambiguous displacement reconstruction, even in the non-linear regime where the voltage is multivalued. The derivative structure further clarifies the contribution of geometric coupling, noise-dominated central behaviour, and envelope decay.
This study establishes, for the first time, a systematic framework for modelling, interpreting, and utilising the full-range behaviour of LVDTs. It offers guidance for systems that undergo large quasi-static excursions, overload recovery, or long-range alignment.\\

\paragraph{keywords:} {Linear Variable Differential Transformer (LVDT); full-stroke sensing; non-linear response; finite element analysis (FEMM); analytical modelling; systematic uncertainties; seismic isolation systems; ETpathfinder; high-precision instrumentation}
\end{abstract}
%------------------------------------------------------------------------------------------------------- 

%=========================================================================================================
\section{Introduction}
Linear Variable Differential Transformers (LVDTs) are widely employed as high-precision, contactless displacement sensors in applications requiring stable, low-noise position readout, for instance robotics and automotive controls \cite{commercialLVDT, Jefriyanto_2020}. Their robustness, vacuum compatibility, and negligible mechanical coupling make them integral to the seismic isolation systems of gravitational-wave detectors such as LIGO \cite{advligo-cite}, Virgo \cite{advvirgo-cite}, KAGRA \cite{kagra-cite}, and the Einstein Telescope Pathfinder (ETpf) \cite{etpf-sensitivity}. In these infrastructures, LVDTs also work in conjunction with voice-coil (VC) actuators to provide co-located sensing and control of suspension stages.

Conventional descriptions of LVDT performance typically emphasise the linear operating range. This linear region is central to precision control, and as such, most design studies, optimisation efforts, parameter variation effects on response and analytical models have been focussed and limited to the small, linear displacements \cite{simulation-paper, Mociran_Gliga_2023, shar-model}. Outside the linear region, the relation between displacement and differential LVDT output voltage departs from a simple proportional law, as confirmed by both finite-element simulations and experimental measurements, which are discussed in the upcoming sections. These non-linearities are not a minor detail: they delimit the usable and safe operating range and directly influence how reliably the LVDT can be employed in feedback control loops, where the full sensing chain implicitly assumes a well-characterised and predictable response. If these assumptions are violated, excess deviation from linearity propagates into calibration errors, degraded control performance, and, incorrect interpretation of the measured motion. A complete understanding of full-range response is therefore essential for accurate displacement reconstruction, robust actuator–sensor decoupling, and effective long-range alignment control. Despite its relevance to precision interferometry and other high-dynamic-range applications, systematic characterisation of the full-range behaviour of LVDTs remained largely unexplored.

This paper presents a detailed modelling framework to analyse LVDT behaviour across the entire mechanical stroke, from the central linear region to the strongly non-linear extremes of travel. Using finite-element modelling with FEMM \cite{femm-site} and its Python interface (pyFEMM) \cite{pyfemm-manual}, we quantify the evolution of magnetic flux linkage, induced voltage, and sensitivity. The simulations incorporate realistic boundary conditions, material properties, coil geometries representative of those used in gravitational-wave detector suspensions. The aim of this study is to move beyond the conventional treatment of LVDTs as just linear sensors and establish a comprehensive description of their full-range displacement response. 

% describe structure of paper
This paper is structured as follows: \autoref{sec:principle} introduces the working principle of the LVDTs along with mathematical expressions, and \autoref{sec:model-def} describes the simulation framework implemented using pyFEMM, together with the geometrical assumptions and modelling procedure followed by the simulation results explaining the transition from linear to non linear regions. \autoref{sec:measurements} presents the experimental results and the analytical expression developed along with the reliability of the model. The final section \autoref{sec:initial conditions} discusses the role of physically informed initial conditions, and the stability of the obtained parameters considering the systematic uncertainties.\\ 
Throughout this paper, the terms ‘primary coil’ and ‘inner coil’ are used interchangeably, as are ‘secondary coil’ and ‘outer coil’.

\subsection{Operating principle} \label{sec:principle}
\begin{figure}[H]
\subfloat[\centering]{\includegraphics[width=0.45\linewidth]{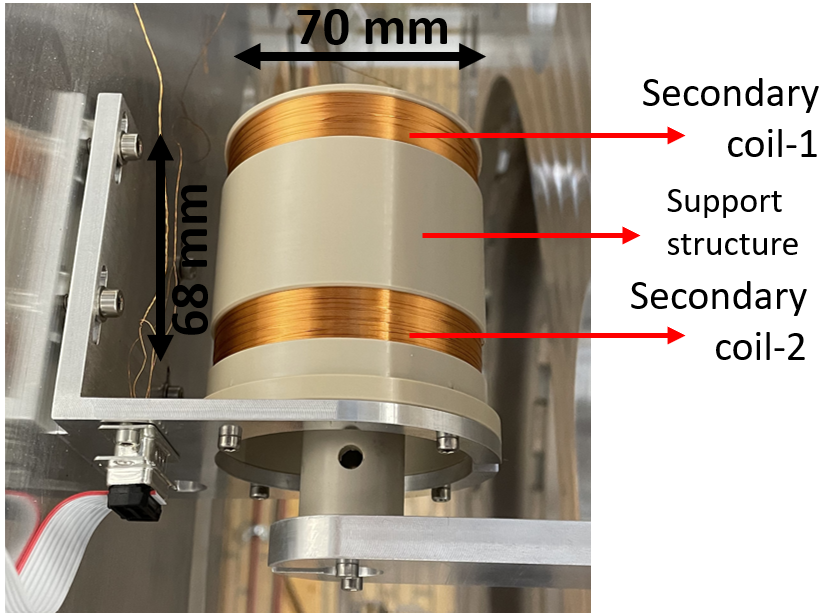}}
\subfloat[\centering]{\includegraphics[width=0.43\linewidth]{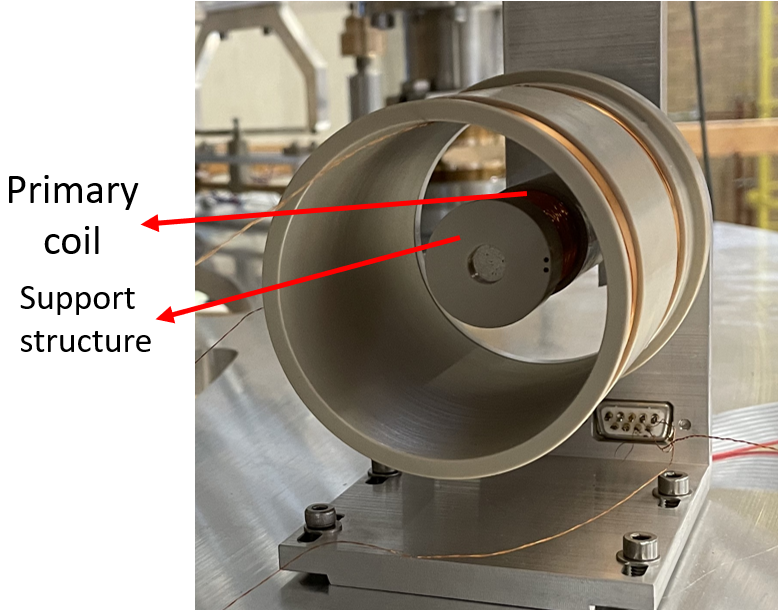}}
\caption{Example of an LVDT installed in a seismic isolation stage of ETpathfinder~\cite{ETpathfinderTDR}. (\textbf{a}) Alignment of the moving primary coil relative to the fixed counter-wound secondary coils. (\textbf{b}) A view of the primary coil, showing the difference in primary and secondary coil radii to allow for residual transverse motion in the suspension. Both views illustrate the efficient design that enables the position sensing.}
\label{fig:lvdt_vc}
\end{figure}

The working principle of LVDTs is based on mutual induction. LVDTs consist of three helical coils, similar to solenoids. One of these coils, called the primary coil, is wound over the support structure which is attached to the object whose relative position change is measured and the other two coils, called the secondary coils, are connected in series and are winded oppositely over another support structure attached to a stable reference structure with the same central axis as of the primary coil, shown in \autoref{fig:lvdt_vc}. 

\begin{figure}[H]
    \centering
    \includegraphics[height = 6cm]{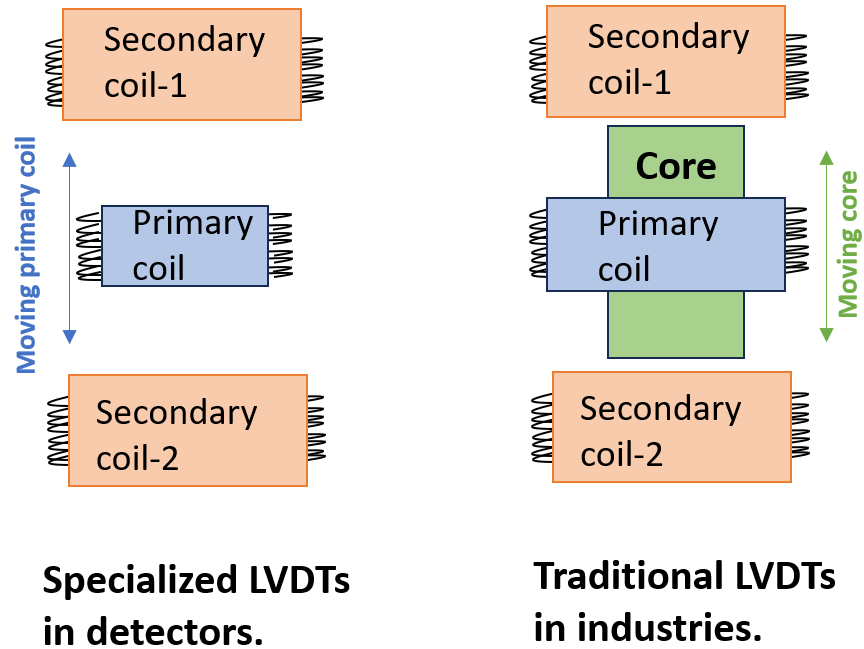}
    \caption{Comparison between moving-primary and conventional core-type LVDTs. Left: moving-primary LVDT configuration used in gravitational-wave suspension systems, where the excited primary coil moves relative to the fixed secondary coils. Right: conventional industrial/commercial LVDT, in which a movable ferromagnetic core modulates the coupling between a stationary primary coil and the secondary coils.}
    \label{fig:lvdt1}
\end{figure}

For traditional LVDTs or so called commercial LVDTs used in industrial grade applications, the moving primary coil is replaced by a moving ferromagnetic core with primary coil having the same radius as of secondary and being a stable reference as shown in \autoref{fig:lvdt1}. Through out this thesis, the terms 'primary coil' and 'inner coil' are used interchangeably, as are 'secondary coil' and 'outer coil'.
Exciting the primary coil with a high frequency sinusoidal signal generates an oscillating magnetic field, which induces a voltage in both the outer coils. As the secondary coils are counter wound the difference between the induced voltages, $\rm V_{\rm difference} = V_{\rm secondary1} - V_{\rm secondary2}$, vanishes if the induction on both coils is exactly the same. This happens only when the source of the induction (inner coil) is at the centre of the outer coils. A small change in the position of the inner coil results in an unequal induced voltage on both the outer coils and hence a non vanishing voltage difference is observed. Detecting this signal serves as a measurement for the position change of the inner coil, i.e, object.
This magnetic field generated due to a sinusoidal excitation of primary coil can be described using the relation \cite{griffiths}:
\begin{equation}
\rm \mathbf{B}(z)=N \frac{\mu_0 I(t) R^2}{2\left(z^2+R^2\right)^{3 / 2}} \mathbf{e}_z.
\end{equation}
The law establishes a relationship between the generated magnetic field, $\rm B(z)$, created by the coil along its symmetric axis where z is the distance from the coil’s centre, containing $\rm N$ number of turns with coil radius $R$ with a current $\rm I(t)$ running through the windings having the vacuum permeability $\rm \mu_o$. Given an (oscillating) input current, $\rm I(t) = I_0\sin(\omega t)$, the magnetic field generated by the equation oscillates at the same frequency. The magnetic field produced by the inner coil (or in a way oscillating magnet), $\rm B(z,t) = B(z)\sin(\omega t)$, subsequently permeates the outer coils. Using Faraday's law \cite{kinsler2020faraday}, the flux generated leads to an induced potential difference $\mathcal{E}$: 
\begin{equation}
\mathcal{E}=-\frac{d \Phi_B(t)}{d t}
\end{equation}
across the outer coils $V_{\rm secondary1}$ and $V_{\rm secondary2}$ due to varying magnetic flux $\Phi_B(t)$. The relation between magnetic flux and magnetic field is given as:
\begin{equation}
\rm \Phi_B(z, t)=\int_{\Sigma} \mathbf{B}(z, t) \cdot d \mathbf{S},
\end{equation}
with $\Sigma$ being the boundary of the outer coil and $d \mathbf{S}$ denoting the infinitesimal surface element perpendicular to the integration surface. 

The amplitude of the induced voltage reaches its maximum value when the centre of the inner coil aligns with the centre of one of the outer coils. The secondary coils produce induced voltages $\rm V_{secondary1}$ and $\rm V_{secondary}$, and the differential output, 
\begin{equation}
    \rm V_{diff} = V_{secondary1} - V_{secondary2},
\end{equation}
changes sign at the centre and is approximately linear for small displacements. This linearity is achieved due to a Maxwell Pair (MP) configuration \cite{tariq2002linear}. This consists of 2 counter winded coils each with radius \textbf{R}, separated by a distance \textbf{d} creating opposite magnetic fields at the centre of the coils. Deviations from linearity arise when the magnetic coupling ceases to scale proportionally with position. 

\section{Model definition} \label{sec:model-def}
\begin{figure}[H]
\centering
\subfloat[\centering]{\includegraphics[width=0.33\linewidth]{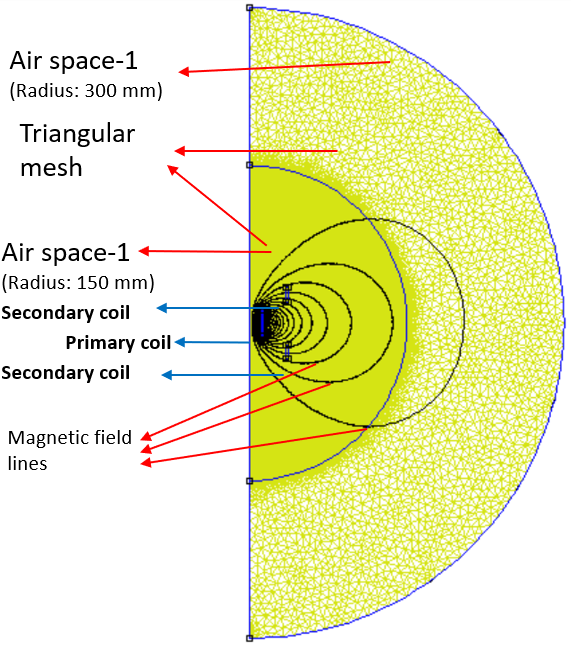}}
\subfloat[\centering]{\includegraphics[width=0.32\linewidth]{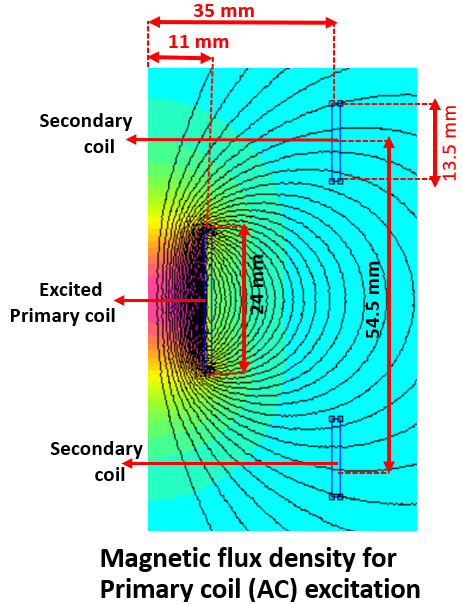}}
\caption{Axisymmetric FEMM simulation results of the LVDT. (\textbf{a}) An overview of the implemented model indicating the airspace volumes and different mesh sizes; (\textbf{b})~LVDT sensing mode with the primary coil excited at 10~kHz. Colour shading indicates magnetic flux density and the contour lines show the field lines linking primary and secondary coils.}
\label{fig:LVDT femm_model}
\end{figure}

The finite element methods are used to model LVDTs starting from early 90s \cite{sykulski}. \texttt{FEMM} (Finite Element Method Magnetics) is an open software \cite{femm-site} designed to solve electromagnetic problems in two-dimensional planar and axisymmetric domains. pyFEMM, a \texttt{Python} extension to perform finite element method analysis using FEMMs, is used throughout the project \cite{pyfemm-manual}. 
\autoref{fig:LVDT femm_model} illustrates the modelling of LVDTs using FEMM. The simulation domain consists of two concentric circular regions representing the ambient air space:  
\begin{itemize}   
    \item \textbf{Airspace-1 (A1)}: region-1 with a radius of 300 mm enclosing the A2 and LVDT. (The space outside A1 is ignored and left un-simulated as the magnetic field is negligible beyond this boundary).
    \item \textbf{Airspace-2 (A2)}: region-2 with a radius of 150 mm surrounding the LVDT with a finer mesh. 
\end{itemize}
The following are the assumptions considered in the model:
\begin{itemize}
    \item LVDT and VC model - Axisymmetric r-z plane where r is the radial and z is the vertical coordinate.
    \item Mesh style - Triangular blocks.
    \item Mesh size - 0.5 (for A2), auto (for A1).
\end{itemize}
Dirichlet boundary conditions are imposed at the airspace with a null vector potential (to prevent the magnetic flux from entering the boundary). 
    For LVDTs, the primary coil was excited with a 10 kHz\footnote{The electronics used in the test bench are optimised for a 10 kHz signal.} sinusoid excitation. The position of the primary coil was varied relative to that of the secondary coil, and the corresponding response was measured at discrete intervals. These simulations generate a series of data points that represent the output signal at different positions. The validity of FEMM simulations had already been verified with theoretical models \cite{femm_validity} and has been used to model the LVDTs in various other works \cite{sykulski, sharfadi, young-model}.

\subsection{Simulation} \label{sec:non linearity-theory}
Throughout this work, we use an LVDT design of the type employed in the ETpf suspension system as a concrete and experimentally accessible example. All simulations and measurement results shown in this and the following sections are based on this specific design\footnote{The adopted dimensions are as follows: inner and outer coil radii: 11 mm and 35 mm; inner and outer coil heights: 24 mm and 13.5 mm; number of wire layers in the inner and outer coils (using 32 AWG wire): 6 and 7, respectively; magnet radius and length: 5 mm and 40 mm; and outer coil separation: 54.5 mm.}. 
\begin{figure}[H]
    \centering
    \includegraphics[width=0.6\linewidth]{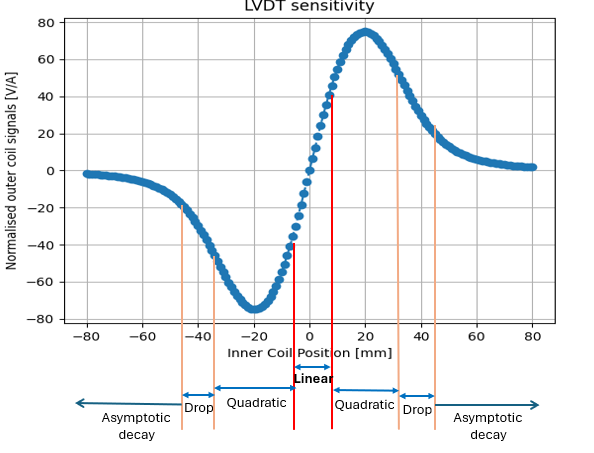}
    \caption[Full displacement range simulation of a Type A LVDT]{Simulation showing the full displacement range of an LVDT. The response is linear near the centre, transitions into a quadratic curvature region, and finally decays asymptotically towards zero.}
    \label{fig:lvdt full range}
\end{figure}
\autoref{fig:lvdt full range} shows the response of an LVDT over its full range until the response vanishes. The response is symmetric about the origin. It remains linear within a certain range, characterised by a slope $\rm k$, which has been the main focus of LVDT analysis in the literature. Beyond this range, the response transitions into a quadratic regime with coefficients $\rm a$ and $\rm b$. This transition occurs as the primary coil enters the plane of one of the secondary coils, say, the upper secondary coil, and begins to pass through it. At a specific position within this region, when the centre of the primary coil aligns with that of the upper secondary coil, the system reaches its peak response. This alignment maximises magnetic coupling and flux linkage, resulting in the highest induced voltage in the upper secondary coil. As the primary coil continues to move beyond this alignment point, the response gradually decreases, reflecting the reduction in induction as the coil exits the secondary region. This marks the onset of the drop-off region, where magnetic interaction weakens substantially. Eventually, for large displacements, the primary coil moves far from both secondary coils, and the induced voltage asymptotically approaches zero. In this regime, the influence of the primary coil becomes negligible, and the response can be modelled using a suitable decay function, such as an exponential or rational form. Combining these observations, the overall response, $\rm f(x)$, can be expressed as a piecewise equation (where $\rm x$ represents the primary-coil displacement):
\begin{equation} \label{eq:piecewise}
    \rm f(x)= \begin{cases}k x, & |x| \leq x_1 \quad \text { (Linear Range) } \\ a\left(x-x_1\right)^2+b, & x_1<|x| \leq x_2 \quad \text { (Quadratic Non-Linear Range) } \\ d e^{-|x| / p}, & x_2<|x| \leq x_3 \quad \text { (Drop-Off Region) } \\ \frac{c}{|x|^n}, & |x|>x_3 \quad \text { (Asymptotic Decay) }\end{cases}
\end{equation}
For $\rm x < 0$, $\rm f(x) = -f(|x|)$ due to symmetry.\\ 
\subsubsection*{Approximation method}
Deriving a single expression to explain the entire response of an LVDT, instead of using piecewise functions shown in \autoref{eq:piecewise}, requires solving the coupled integrals as functions of coil position, geometry, and field distributions explained in \autoref{sec:principle}. Such equations are complicated to solve and difficult even to find the existence of a solution. \\
Instead, a reasonable approximation, using the measured data, can be achieved by smoothly blending the different regions using polynomial, exponential, or sigmoid functions. A careful inspection of the simulated response, shown in \autoref{fig:lvdt full range}, reveals several robust structural features: 
\begin{itemize}
    \item Linearity around the centre, characterised by a constant first derivative.
    \item Smooth quadratic curvature that develops as the primary coil approaches the secondary windings.
    \item A well-defined maximum (or minimum) where magnetic coupling is strongest.
    \item A monotonic drop-off region, where the induced voltage rapidly decreases.
    \item A slowly asymptotic decay, consistent with the diminishing flux linkage at large separation.
    \item Odd symmetry $\rm f(-x) = -f(x)$.
\end{itemize}
After testing several families of functions, polynomial, rational, piecewise spline, and trigonometric forms,the following function emerged as the simplest expression, capturing all these behaviours within a single continuous function:
\begin{equation} \label{eq:lvdt-eq}
    \rm f(x) = Ae^{-Bx^{2}}\sin(Cx)+Dxe^{-Ex^{2}}.
\end{equation}
The first term accounts for oscillations observed in the response, with the amplitude modulated by $\rm e^{-Bx^{2}}$ and gradually diminishing for large $\rm x$. The second term captures the linear and quadratic behaviour near the central region and ensures a smooth asymptotic decay. It anchors the linear response at small $x$ and guides the transition into the quadratic region. Note that the expression \autoref{eq:lvdt-eq} is not the analytical solution of the LVDT electromagnetic problem. \\ 
\subsubsection*{Linear range}
For small x, the exponential and sine terms can be approximated using a Taylor expansion
\begin{equation}
    \rm e^{-Bx^{2}} \simeq 1 - Bx^{2} + \mathcal{O}(x^{4}), e^{-Ex^{2}} \simeq 1 - Ex^{2} + \mathcal{O}(x^{4})
\end{equation}
\begin{equation}
    \rm \sin(Cx) \simeq Cx - \frac{(Cx)^{3}}{6}+\mathcal{O}(x^{5}).
\end{equation}
Hence, the response $\rm f(x) \sim (AC + D)x$ is approximately linear.\\
\subsubsection*{Quadratic range}
As $x$ increases, the exponential term $\rm e^{-Ex^{2}}$ begins to play a significant role, leading to:
\begin{equation}
    \rm f(x) \sim Ae^{-Bx^{2}}\sin(Cx)+Dx(1-Ex^{2})
\end{equation}
Here, the second term dominates, making the response approximately quadratic:
\begin{equation}
    \rm f(x) \sim Dx - DEx^{3}
\end{equation}
\subsubsection*{Drop region}
The sinusoidal term in $\rm f(x)$ introduces oscillations into the decaying portion of the response, whose amplitudes decrease progressively with the diminishing factor $\rm e^{-Bx^{2}}$.\\
\subsubsection*{Asymptotic region}
For large $\rm x$, both exponential terms, $\rm e^{-Bx^{2}}$ and $\rm e^{-Ex^{2}}$, dominate and decay rapidly. Consequently, the response approaches zero:
\begin{equation}
    f(x) \simeq 0 \ \rm for \ |x|\gg0 
\end{equation}
\section{Measurement} \label{sec:measurements}
\begin{figure}[H]
    \centering
    \includegraphics[width=0.6\linewidth]{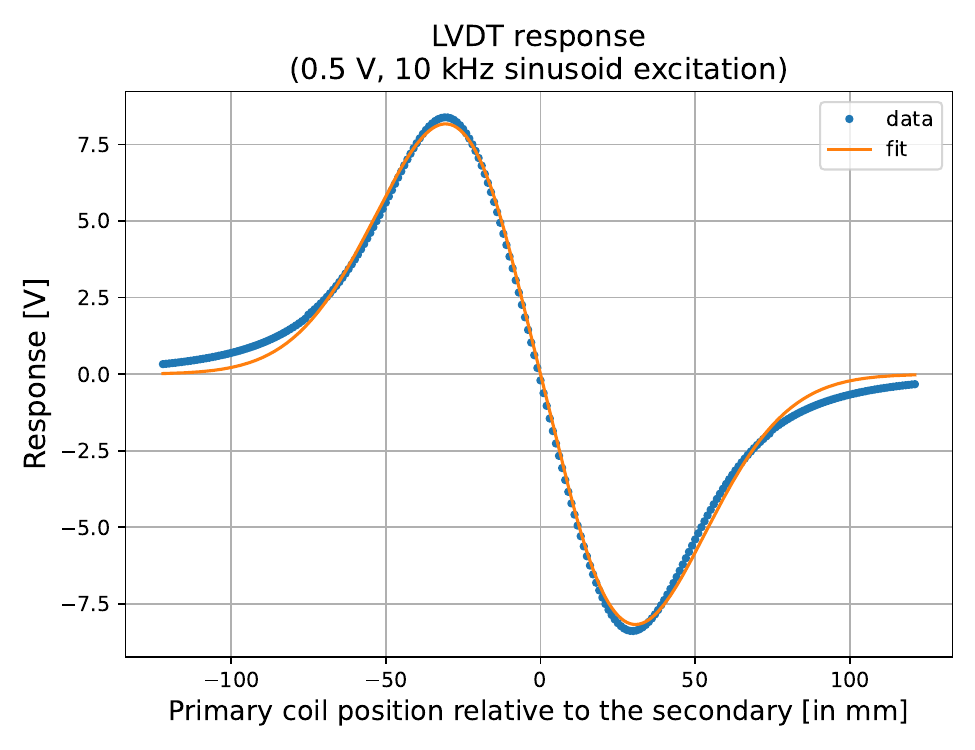}
    \caption{Fitted model compared to the measured response, combining polynomial and exponential components to capture the transition from linearity to asymptotic decay. }
    \label{fig:lvdt full range fitl}
\end{figure}

\begin{figure}[H]
    \centering
    \includegraphics[width=0.6\linewidth]{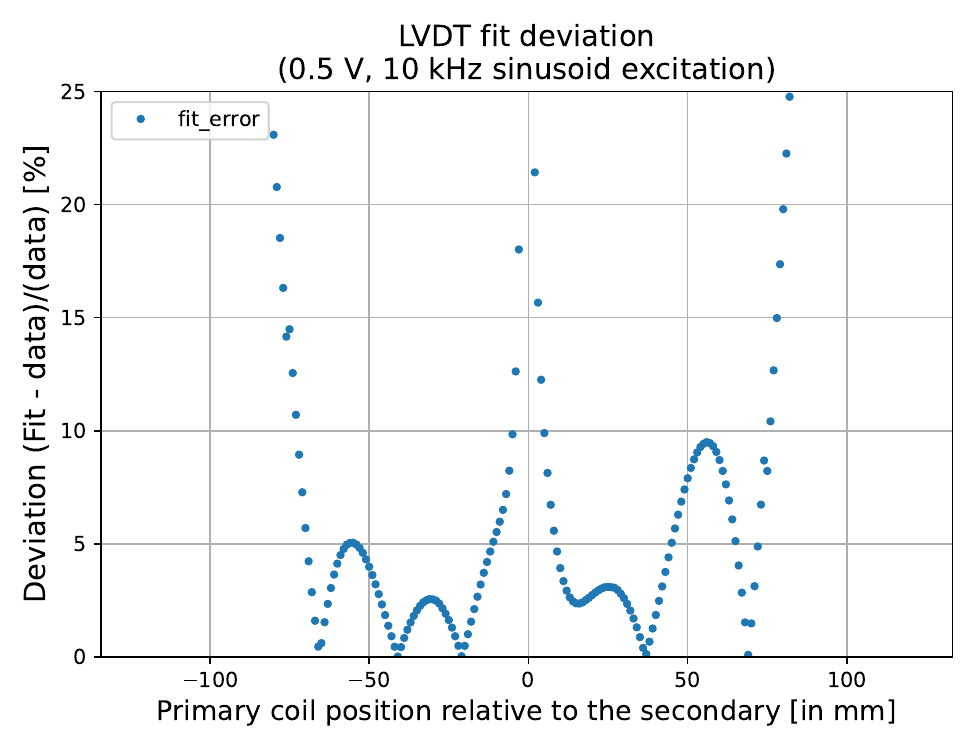}
    \caption{The deviation $\frac{(\rm fit \ data \ - \ original \ data)}{\rm original \ data}$ across the full displacement range. The deviation remains below approximately 5\% over a displacement range of about 115 mm (from -70 mm to 45 mm). Larger deviations occur in the asymptotic regions where the signal approaches zero and the fit becomes sensitive to small parameter variations. The apparent divergence near the centre arises from the normalisation in the relative error definition, where the signal amplitude approaches zero, producing a mathematical artefact.}
    \label{fig:lvdt full range fit dev}
\end{figure}

The measured response from the LVDT is fitted using the \autoref{eq:lvdt-eq}  and is shown in \autoref{fig:lvdt full range fitl}. The initial guess for the curve fit is [1, 1, 1, 1, 1] corresponding to [A, B, C, D, E] in \autoref{eq:lvdt-eq}.
\autoref{fig:lvdt full range fitl} demonstrates that the unified compact expression, \autoref{eq:lvdt-eq}, estimates the response accurately, capturing the transition from the central linear regime to the quadratic and asymptotic domains. The values of the parameters A, B, C, D, and E obtained for three repetitions is presented in \autoref{tab:syst}:
\begin{table}[H]
    \centering
    \begin{tabular}{|c|c|c|c|c|c|}
    \hline
         \small{Parameter}&A&B&C&D&E  \\
         \hline
         \small{Value}&\small{-64.80}$\pm$4.44&\small{$(5.000 \pm 0.010)\times10^{-4}$}&\small{0.0068$\pm$0.0004}&\small{$(3.300 \pm 0.008)\times10^{-2}$}&\small{$(1.100 \pm 0.001)\times10^{-2}$}\\
         \hline
    \end{tabular}
    \caption{The values of the constants [A, B, C, D, E] obtained using arbitrary initial conditions as [1, 1, 1, 1, 1].}
    \label{tab:syst}
\end{table}
The relative deviation, difference between data and fit compared to the data,  remain less than 5\% for about 115 mm stroke (from -70 mm to 45 mm), as shown in \autoref{fig:lvdt full range fit dev}. Within this interval, the unified compact model reproduces the measured response with good accuracy, indicating that the functional form captures the dominant behaviour of the LVDT over most of its operational stroke. Outside this region, particularly towards the edges of the measured range, the deviation increases. This is due to two reasons: (i)As seen in \autoref{fig:lvdt full range fitl}, the arbitrary initial conditions used in this fit do not fully capture the exact curvature of the response in the asymptotic region, resulting in larger discrepancies between the model and the measured data. (ii)The asymptotic decay of the LVDT signal, where the voltage approaches zero and the fit becomes more sensitive to small variations in the parameters.

A second apparent increase in the relative deviation occurs near the centre. In this region the signal amplitude itself approaches zero, which leads to a mathematical singularity in the relative error definition. Since the relative deviation is computed as $\frac{|(\rm fit \ data \ - \ original \ data)|}{\rm original \ data}$, the denominator vanishes near the centre, producing a $\frac{0}{0}$ type expression. This causes the relative deviation to spike to very large values or formally diverge, even though the absolute difference between the data and the fit remains small. So, this is a mathematical artefact of the normalisation, rather than a physical limitation of the model or the sensor.

The deviations observed in the asymptotic region can be significantly reduced by selecting more appropriate initial parameter estimates for the fitting procedure which will be discussed in \autoref{sec:initial conditions}.

While the unified expression, \autoref{eq:lvdt-eq}, reproduces the full response curve, it is not inherently invertible in the non-linear regime. In the quadratic region, and particularly around the peak response, a given voltage level corresponds to two distinct mechanical positions, one on the rising branch (before the peak) and one on the falling branch (after the peak). Recovering the correct position from the measured voltage therefore requires additional information. The first derivative of the response provides the necessary information to resolve this ambiguity. This reveals the monotonicity and curvature of the response, allowing unique identification of the operating region.
\subsection{First derivative}
\begin{figure}[H]
    \centering
    \includegraphics[width=0.5\linewidth]{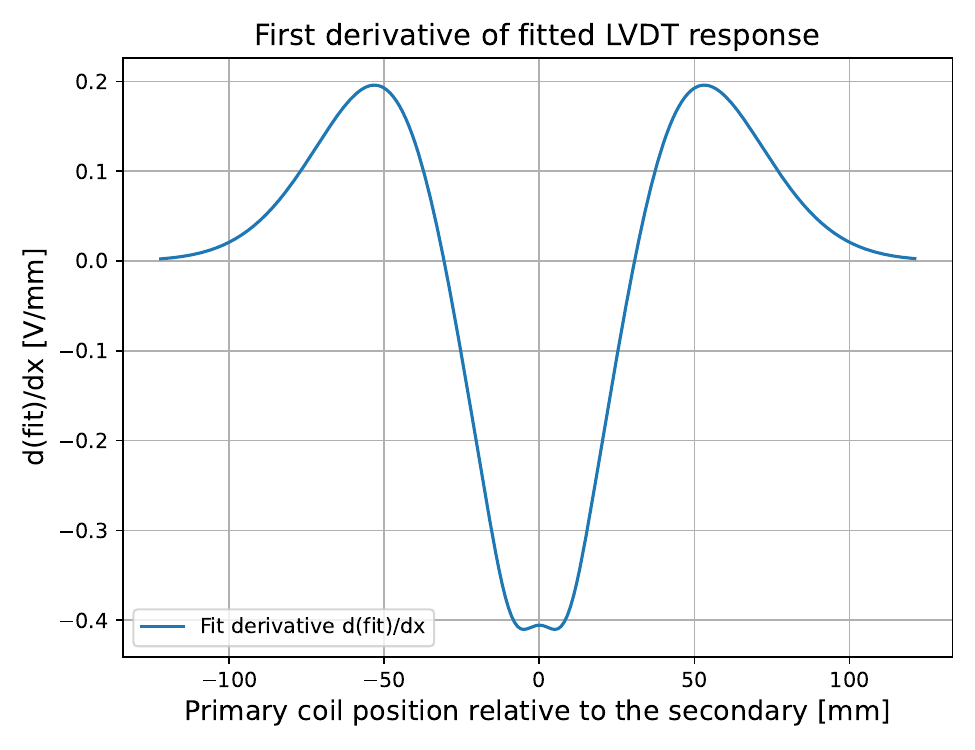}
    \caption[First derivative of the unified full-range fit]{First derivative of the unified compact fit, $\rm f'(x)$, showing an approximately constant slope in the central linear region, followed by a gradual change in the non-linear regime. The derivative crosses zero at $x \approx \pm 30~\mathrm{mm}$, coinciding with the maximum and minimum of the response. The sign of $\rm f'(x)$ discriminates between the rising and falling branches of the LVDT response, enabling unambiguous position retrieval in regions where the same voltage occurs at two different displacements.}
    \label{fig:fullrange-first-deriv}
\end{figure}

Differentiating \autoref{eq:lvdt-eq} yields:
\begin{equation} \label{eq:lvdt-first}
    \rm f^{'}(x) = Ae^{-Bx^{2}}[C\cos(Cx)-2Bx\sin(Cx)]+Dxe^{-Ex^{2}}(1-2Ex^2).
\end{equation}
The corresponding curve is shown in \autoref{fig:fullrange-first-deriv}. Around the  centre $\left| x \right| \leq 5$ mm, $\rm f^{'}(x)$ is nearly constant, with an average slope of $\rm \left| \frac{df}{dx} \right| \simeq 0.4$. This confirms the expected linear behaviour in this central operating range. As $\left| x \right|$ increases, the derivative deviates from this constant value, signalling the onset of non-linearity. 

The prominent features of \autoref{fig:fullrange-first-deriv} are:
\begin{itemize}
    \item two symmetric zero-crossings at $\left| x \right| \sim \pm 30$ mm, coinciding with the locations of the global maximum and minimum of $\rm f(x)$
    \item a negative minimum around $\rm \left|x\right| \simeq 0$, corresponding to the steepest part of the response where the output changes sign.
\end{itemize}
Physically, the sign of $\rm f(x)$ encodes whether the LVDT operates on the rising or falling branch of the non-linear response. In the region where a given voltage occurs at two different positions (one on each side of the peak), the measured derivative can be used to resolve this ambiguity:
\begin{itemize}
    \item $\rm f^{'}(x) > 0$ indicates that the position is on the falling branch (or after the peak),
    \item $\rm f^{'}(x) < 0$ indicates that the position is on the rising branch (or before the peak).
\end{itemize}
Hence the first derivative provides a robust monotonicity marker for position retrieval in the non-linear regime.
\subsection{Second derivative}
\begin{figure}[H]
    \centering
    \includegraphics[width=0.5\linewidth]{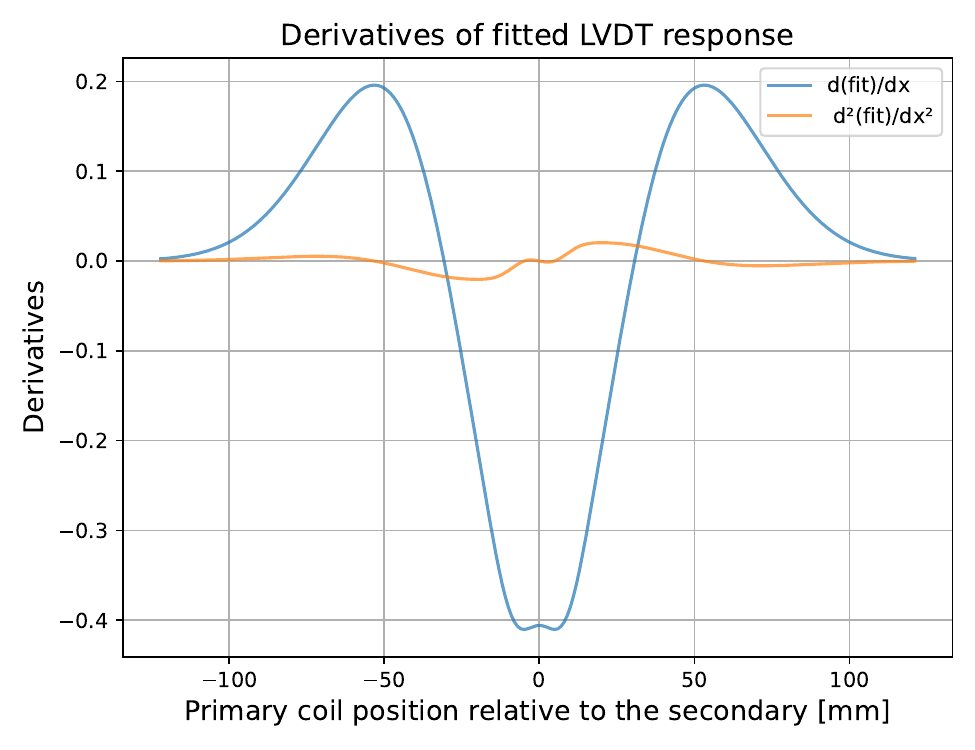}
    \caption[Derivatives of the unified full-range fit]{Overlay of $\rm f'(x) \ and \ f''(x)$, showing monotonicity and curvature. The curvature of the second derivative is close to zero in the central linear region, becomes positive when approaching each peak (convex rising branch), and negative after the peak (concave falling branch), with inflection points where $f''(x)=0$. Together with $f'(x)$, this curvature information allows robust classification of the operating regime and supports unique displacement reconstruction across the full stroke.}
    \label{fig:fullrange-derivs}
\end{figure}

A second differentiation of \autoref{eq:lvdt-eq} gives the curvature
\begin{equation}
    \rm f^{''}(x) = \frac{df^{'}(x)}{dx},
\end{equation}
which was evaluated numerically from the fitted parameters. The resulting curve is compared with \autoref{fig:fullrange-first-deriv} and is shown in \autoref{fig:fullrange-derivs}. 

% \begin{figure}[H]
%     \centering
%     \includegraphics[width=0.5\linewidth]{images/fullrange_exp_der2.pdf}
%     \caption[Second derivative of the unified full-range fit]{Second derivative of the unified compact fit, $f''(x)$, highlighting the curvature of the LVDT response. The curvature is close to zero in the central linear region, becomes positive when approaching each peak (convex rising branch), and negative after the peak (concave falling branch), with inflection points where $f''(x)=0$. Together with $f'(x)$, this curvature information allows robust classification of the operating regime and supports unique displacement reconstruction across the full stroke.}
%     \label{fig:fullrange-second-deriv}
% \end{figure}

Near the centre, the curvature is close to zero, as expected for an almost perfectly linear response. Moving away from the centre, $\rm f^{''}(x)$ becomes positive on the rising side of each hump and negative on the falling side. The point where $\rm f^{''}(x)=0$  mark inflection points that separate these regimes. This curvature information is particularly useful when the first derivative alone becomes small, for example close to the peak of the response. In that case:
\begin{itemize}
    \item $\rm f^{'}(x) \simeq 0 \ and \  f^{''}(x)>0$ indicate that the position is on the convex, positive part of the response;
    \item $\rm f^{'}(x) \simeq 0 \ and \  f^{''}(x)<0$ indicate that the position is on the negative part of the response.
\end{itemize}
Thus, combining the signs of $\rm f, f^{'}, f^{''}$ allow unambiguous classification of the operating region (central linear, pre-peak, post-peak, or asymptotic), and thus unique reconstruction of the displacement.
\subsection{Demodulation}
If the LVDT output is demodulated, the process recovers both the amplitude of the secondary voltage and its phase relative to the excitation. Since the induced signal reverses sign when the primary coil crosses the centre, the demodulated output already contains explicit polarity information. Thus, for any point in the response curve, one directly knows whether the displacement is on the positive branch or the negative branch of the LVDT response. This means that, in a practical measurement, the sign of the displacement does not need to be inferred from higher-order derivatives. 

Although the LVDT response is fundamentally a static mapping from (induced voltage to displacement), the combined information contained in $\rm (f(x), f'(x), f''(x))$ naturally forms a state vector; multi-dimensional representation of the sensor’s state space. In this sense, each component in the state vector contributes independent information that resolves ambiguities which cannot be distinguished from the output alone. While the voltage $\rm f(x)$ may be identical at two different positions in the non-linear regime, the first derivative identifies whether the sensor lies on the rising or falling branch, and the second derivative distinguishes pre-peak from post-peak curvature. Together, these derivatives uniquely determine the “state” of the system across the full displacement range, enabling unambiguous position reconstruction.
\subsection{Reliability of the model}
 As the compact expression introduced in \autoref{eq:lvdt-eq} is not obtained from first-principles derived directly from Maxwell’s equations, an important question is therefore whether this behaviour is unique to the specific LVDT studied in this work or whether they arise generally. The \autoref{eq:lvdt-eq} captures the nature of any physical system that exhibits linear, quadratic, drop-off and asymptotic behaviours sequentially. Considering the LVDTs, the physical origin of each regime can be understood as follows:

\begin{itemize}
    \item \textbf{Linear region (near centre): }When the primary coil is centred between the two secondary coils, this linear nature arises because the field distribution is nearly symmetric and the voltage difference grows proportionally to the small displacement.
    \item \textbf{Quadratic region: }As the coil moves further, the induced voltage no longer scales linearly. This curvature arises from non uniform changes in flux linkage.
    \item \textbf{Drop-off region: }Once the moving coil passes the central region of a secondary, the induced voltage begins to decline. The coupling reduces because the primary is partially outside the secondary winding.
    \item \textbf{Asymptotic region: }For large $\left| x \right|$, the primary is far from both the secondary coils and the magnetic coupling becomes negligible. FEMM simulations confirm that the induced voltage asymptotically approaches zero. This naturally matches the behaviour of functions such as $\rm e^{-Bx^2}, \frac{1}{x^n}$, or other decaying envelopes.
\end{itemize}
Thus, any function that captures the combination of linear, quadratic, peak, drop-off, and asymptotic regions will describe a broad class of LVDTs, not just the specific one measured. Although fitted using data from a single LVDT, the \autoref{eq:lvdt-eq} is not specific to that one sensor. Its suitability is rooted in universal physical behaviour common to all LVDTs, regardless of coil dimensions or construction.
\section{Initial conditions} \label{sec:initial conditions}
In the previous sections, the initial assumptions for fitting the five parameters,A, B, C, D, and E , were taken to be unity. It should be noted that such arbitrary initialisations do not necessarily guarantee convergence to an optimal fit always. For a more accurate and stable convergence, the choice of realistic initial conditions is of considerable importance. By utilising simple geometric parameters extracted from the measured response, the five parameters can be initialised with values that are physically meaningful. This approach not only improves the quality of the fit but also reduces computational time. 

A rigorous and analytical framework for determining optimal initial conditions constitutes a separate line of investigation that warrants further study. Nevertheless, a preliminary attempt has been undertaken here to estimate suitable initial values, which may be explored in greater detail in future work. The following steps outline the procedure:
\subsubsection*{Initial amplitude ($\rm A_{0}$)}
The dominant oscillatory contribution in \autoref{eq:lvdt-eq} is the term, $\rm Ae^{-Bx^2}\sin(Cx)$. As the measured LVDT response achieves a maximum magnitude of 8.4 V (see \autoref{fig:lvdt full range fitl}), the amplitude parameter was initially set to a value of the same order, but slightly lower to avoid overestimation:
\begin{equation}
    \rm A_0 = 8.
\end{equation}
\subsubsection*{Envelope decay parameters ($\rm B_0, E_0$)}
Both exponential terms $\rm e^{-Bx^2}$ and $\rm e^{-Ex^2}$. control how rapidly the response decays away from the central region. Empirically, the measured LVDT signal falls to nearly zero for $|x|\geq90-100$ mm. A Gaussian envelope satisfying $\rm e^{-Bx^2} \simeq 0.01$ for $\rm x\sim 100 \implies B \sim 10^{-4}$. Thus both decay parameters can be initialized as
\begin{equation}
    \rm B_0, E_0 \simeq 10^{-4},
\end{equation}
This ensures the model initially exhibits a similar bandwidth to the measured response.
\subsubsection*{Initial oscillation frequency ($\rm C_0$)}
The sinusoidal factor $\rm \sin(Cx)$ captures the mild oscillatory curvature in the non-linear regime. The period of this modulation is
\begin{equation}
    \rm \lambda = \frac{2\pi}{C}.
\end{equation}
\autoref{fig:lvdt full range fitl} shows that the response exhibits one broad hump on each side and no rapid oscillations. Therefore the wavelength must be larger than the stroke. A conservative initial value corresponding to a wavelength of 260 mm ($\rm \sim130 \ mm$ on both the sides), 
\begin{equation}
    \rm C_0 = 0.024, 
\end{equation}
was considered.
\subsubsection*{Initial slope parameter ($\rm D_0$)}
Expanding the model near the centre yields
\begin{equation}
    \rm f(x) \sim (AC+D)x,
\end{equation}
so the initial central slope must match the experimentally observed sensitivity. From a simple linear fit to the measured response for $|x|\leq 5$ mm, the slope is approximately:
\begin{equation}
    \rm k \sim 0.44 \ V/mm.
\end{equation}
Given the initial choices $\rm A_0 = 8$ and $\rm C_0 = 0.02$, the contribution from the first term is $\rm A_0C_0 = 0.16$. Hence the $\rm D$ term is:
\begin{equation}
    \rm D_0 = k - A_0C_0 = 0.44-0.16 = 0.28.
\end{equation}

\begin{figure}[H]
    \centering
    \includegraphics[width=0.5\linewidth]{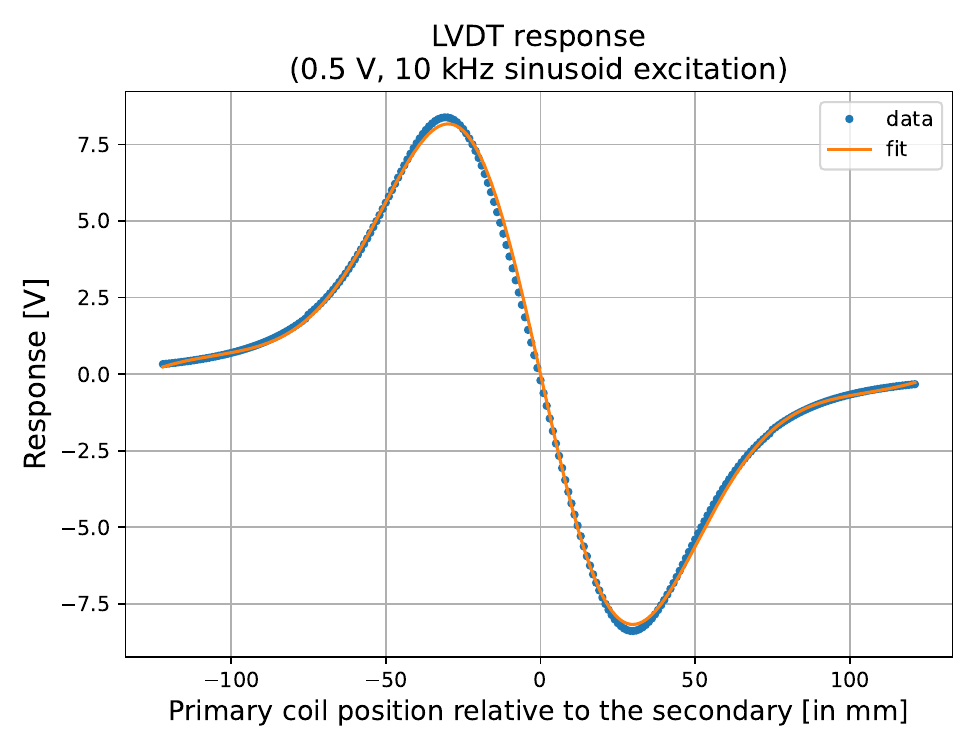}
    \caption{Full-range LVDT response together with the unified compact fit obtained using physically informed initial parameter estimates. Compared to the fit obtained from arbitrary initial conditions, the model reproduces the curvature and asymptotic decay of the response more accurately, leading to improved agreement with the measured data across the entire displacement range.}
    \label{fig:full fit initial guess}
\end{figure}

\begin{figure}[H]
    \centering
    \includegraphics[width=0.5\linewidth]{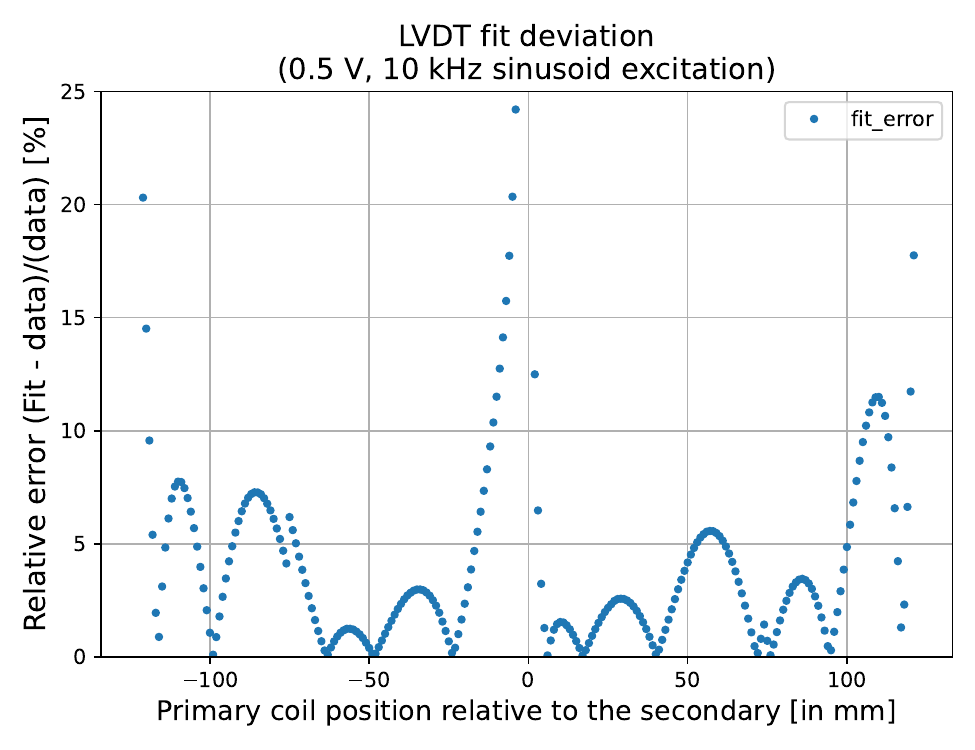}
    \caption{The deviation obtained when \autoref{eq:lvdt-eq} is initialised using a reasonable parameter estimates. Compared to fits obtained using arbitrary initial values, the overall deviation is reduced (except at the region close to the centre), particularly in the non-linear and decay regions. The deviation remain almost symmetric, and bounded below 5\%, demonstrating a more accurate reproduction of the measured LVDT response.}
    \label{fig:dev-initial-guess}
\end{figure}

Therefore, the final initial parameters used for fitting were:
\begin{equation}
    \rm A_0, B_0, C_0, D_0, E_0 = 8, 10^{-4}, 0.024, 0.28, 10^{-4}.
\end{equation}
The LVDT response is fitted with the above initial condition is shown in \autoref{fig:full fit initial guess}.
The fit obtained from the above logically informed initial conditions shows better performance than the arbitrary initial conditions with comparatively less deviation, especially in the asymptotic region, as shown in \autoref{fig:dev-initial-guess}.
 
The values of the parameters A, B, C, D, and E obtained for three repetitions is presented in \autoref{tab:syst2}:
\begin{table}[H]
    \centering
    \begin{tabular}{|c|c|}
    \hline
         Parameter & Value \\
         \hline
         $\rm A$ & $-0.30020 \pm 0.00042$ \\[2pt]

         $\rm B$ & $(-1.1000 \pm 0.0014)\times10^{-4}$ \\[2pt]

         $\rm C$ & $(2.449000 \pm 0.000114)\times10^{-2}$ \\[2pt]

         $\rm D$ & $-0.445960 \pm 0.000036$ \\[2pt]

         $\rm E$ & $(5.80 \pm 0.65)\times10^{-4}$ \\
         \hline
    \end{tabular}
    \caption{Fitted values of the parameters $\rm [A,B,C,D,E]$ obtained using the physically informed initial conditions $[8,\ 10^{-4},\ 0.024,\ 0.28,\ 10^{-4}]$. Uncertainties correspond to the standard deviation obtained from three independent fit repetitions.}
    \label{tab:syst2}
\end{table}
\subsection*{Systematic uncertainties from transverse misalignment}
The parameter values reported in Tab.~\ref{tab:syst2} represent the nominal
best-fit solution obtained under the assumption of perfect axial alignment
between the primary and secondary coils. The corresponding uncertainties are
purely statistical, estimated from the variance over three independent fit
trials. In the experiment, however, a small transverse offset of the primary
coil relative to the secondary pair is unavoidable. To quantify the resulting
systematic effect, we repeated the full-range fit for two extreme cases
corresponding to a $\pm 1~\mathrm{mm}$ transverse offset. The resulting
parameter sets are summarised in Tab.~\ref{tab:syst-off}.
\begin{table}[H]
    \centering
    \begin{tabular}{|c|c|c|c|c|c|c|}
    \hline
        Transversal&&&&&& \\
         offset&Parameter&A&B&C&D&E  \\
         \hline
         +1mm&Value&-0.304355&-0.000106&0.02448&-0.44700&0.000579 \\
         \hline
          -1mm&Value&-0.292820&-0.000100&0.02458&-0.45531&0.000577 \\
         \hline
    \end{tabular}
    \caption{Fitted parameter values $\rm [A,B,C,D,E]$ obtained under deliberate $\pm 1$\,mm transverse offsets of the primary coil relative to the secondary coil pair. These values are used to estimate the alignment-related systematic uncertainties shown in Tab.~\ref{tab:syst}. The asymmetric shifts between the $+1$\,mm and $-1$\,mm cases reflect the non-linear sensitivity of each parameter to transverse misalignment.}

    \label{tab:syst-off}
\end{table}
% These values are compared with the nominal fit values in Table~\ref{tab:syst2}. The mean absolute deviations between the nominal and offset fits provide an estimate of the systematic uncertainty associated with a $\rm ±1$ mm transverse offset:

% These values are larger than the purely statistical fit uncertainties for some parameters, emphasising that alignment tolerances constitute the dominant error source when interpreting the absolute parameter values A, B, C, D, and E. In the following, we therefore focus on physically meaningful combinations of parameters that directly govern the linear and quadratic response regimes.\\

For each parameter $\rm P \in \{A,B,C,D,E\}$, we treat the nominal value in
Tab.~\ref{tab:syst2} as $\rm P_{\rm mean}$ with statistical uncertainty
$\sigma_{\rm stat}$, and construct systematic shifts via
\begin{equation}
\rm \Delta P_{+} = P(+1~\mathrm{mm}) - P_{\rm mean}, \qquad
\rm \Delta P_{-} = P(-1~\mathrm{mm}) - P_{\rm mean}.
\end{equation}
The total upper and lower uncertainties are then obtained by combining the
statistical and systematic contributions in quadrature,
\begin{equation}
\rm \sigma_P^{(+)} = \sqrt{\sigma_{\rm stat}^2 + \Delta P_{+}^2}, \qquad
\rm \sigma_P^{(-)} = \sqrt{\sigma_{\rm stat}^2 + \Delta P_{-}^2}.
\end{equation}
This yields asymmetric uncertainty bounds
$\rm P_{\rm mean} + \sigma_P^{(+)}$ and $\rm P_{\rm mean} - \sigma_P^{(-)}$, reflecting
the fact that the $+1~\mathrm{mm}$ and $-1~\mathrm{mm}$ offsets do not shift the
parameters symmetrically.
\begin{table}[H]
\centering
\begin{tabular}{|c|c|}
\hline
Parameter & Value (mean$^{+\sigma^{(+)}}_{-\sigma^{(-)}}$) \\
\hline
$\rm A$ & $-0.300200^{+0.007396}_{-0.004175}$ \\[2pt]

$\rm B$ & $-0.000110^{+0.000002}_{-0.000001}$ \\[2pt]

$\rm C$ & $0.024490^{+0.000017}_{-0.000005}$ \\[2pt]

$\rm D$ & $-0.445960^{+0.000426}_{-0.001050}$ \\[2pt]

$\rm E$ & $0.000580^{+0.000001}_{-0.000001}$ \\
\hline
\end{tabular}
\caption{Nominal parameter values and asymmetric uncertainties for the unified compact LVDT model. All parameters are expressed using a uniform decimal format with six digits after the decimal point. Uncertainties include statistical contributions from repeated fits and systematic contributions derived from $\pm 1$~mm transverse-offset tests.}
\label{tab:syst-uncertainties}
\end{table}

Although asymmetric uncertainties for each of $\rm \{A,B,C,D,E\}$ is important, the physically most relevant quantities are the combinations that determine the response behaviour in the linear and quadratic regimes.
Expanding the unified model near the central, linear region yields
\begin{equation}
\rm f(x) \simeq (AC + D)\,x, 
\end{equation}
so the effective linear sensitivity is governed by the combination $\rm AC+D$.
Using the nominal parameter values from Tab.~\ref{tab:syst2}, we obtain
\begin{equation}
\rm AC + D = -0.4533.
\end{equation}
Propagating the statistical uncertainties via
\begin{equation}
\rm \sigma_{AC+D,{\rm stat}}^2
= C^2\sigma_A^2 + A^2\sigma_C^2 + \sigma_D^2,
\end{equation}
yields the resulting systematic uncertainties as
\begin{equation}
\sigma_{AC+D}^{(+)} \simeq +0.000463, \qquad
\sigma_{AC+D}^{(-)} \simeq-0.001055,
\end{equation}
and we obtain
\begin{equation}
\rm AC + D =-0.453312^{+0.000463}_{-0.001055}.
\end{equation}

In the quadratic non-linear regime, the dominant contributions arise from
$\rm D$ and the product $\rm DE$, which sets the leading-order curvature for larger
displacements. Applying the same error-propagation procedure gives
\begin{equation}
\rm D =-0.445960^{+0.000426}_{-0.001050},
\end{equation}
and
\begin{equation}
\rm DE = -0.000259^{+0.000001}_{-0.000001}.
\end{equation}
Together, the combinations $\rm (AC+D)$, $\rm D$, and $\rm DE$ provide a concise and
physically interpretable summary of the LVDT's linear and quadratic response,
including both statistical and alignment-related systematic uncertainties.

\subsection*{Experimental setup used for LVDT characterisation}
To experimentally characterise the full-range response of the LVDT used in this study, a precision translation system was implemented to control the displacement of the primary coil relative to the fixed secondary coil. The setup consists of two orthogonal linear stages: a PI M-406 micro-translation stage providing vertical motion with a positioning uncertainty of 0.2 $\rm \mu m$, and a PI VT-80 translation stage enabling horizontal motion with an uncertainty of 0.4 $\rm \mu m$. These stages allow accurate and repeatable positioning of the primary coil across the entire mechanical stroke.

The LVDT response is routed to a custom amplification board and digitised using the modular DaqBox system developed by LAPP Annecy \cite{inst-lapp} for Virgo. The DaqBox is a mezzanine-based acquisition and control platform incorporating an embedded ARM processor with voltage and temperature monitoring. It supports data transfer via 1-Gbps Ethernet and 2.5-Gbps TOLM optical links and can interface with up to two DSPs per mezzanine, providing a stable and low-noise readout environment.

The primary coil is excited with 10 kHz frequency\footnote{Chosen to match the electronics and DAC configurations used in the setup.} and at each discrete displacement step, the digitised secondary-coil signal is fitted to a sinusoidal model
\begin{equation}
    \rm V(t) = A\sin(\omega t + \phi),
\end{equation}
from which the amplitude $\rm A$ is extracted and taken as the LVDT response for that position. Repeating this procedure across the full operational range yields the complete experimental response curve used in this work. A more detailed description of the experimental setup, electronics, and the measurement procedure used in this work is presented in our earlier study \cite{setup-paper}, focusing specifically linear-range characterisation of the LVDTs.

\section{Conclusions}
\label{sec:conclusions}
In this paper we have presented a complete characterisation of the full-range behaviour of a moving-primary LVDT, extending well beyond the conventional linear operating region which was not a primary focus in the existing literature. Using a combination of finite-element modelling, direct measurements, and an analytically compact parametric model, we have shown that the LVDT response contains several distinct dynamical regimes, central linear, quadratic non-linear, peak coupling, drop-off, and asymptotic decay. These regions arise naturally from the electromagnetic geometry and mutual-inductance structure of the device and define the sensor’s usable displacement range. For the ETpathfinder LVDT studied here, these regimes span a total displacement of approximately 250 mm, with the response attaining a maximum amplitude of about 8.3 V near 25 mm from the electrical centre.

The unified compact expression introduced here provides a single, continuous representation of the entire response curve and reproduces the measured data with high accuracy. Its analytical derivatives further allow unique recovery of the displacement in regions where the voltage is multivalued, enabling reliable position reconstruction across the full stroke. A detailed uncertainty analysis shows that while statistical variations in the fitted parameters [A,B,C,D,E] remain small (at the level of $10^{-4}-10^{-6}$), transverse misalignment of the coils introduces systematic shifts at the $10^{-3}$ level. This framework offers a practical, physics-informed alternative to high-order polynomial or spline models, while remaining computationally lightweight and suitable for control-oriented applications.

By explicitly quantifying the non-linear and asymptotic behaviour, this study provides essential guidance for the use of LVDTs in precision engineering, industrial sensing, and advanced suspension systems, where large excursions, recovery after overloads, and accurate calibration across extended ranges are required. 

\section{Acknowledgements}
 The author acknowledges the support of the laboratory and technical facilities at the University of Antwerp, where the experimental measurements presented in this work were carried out. The author also acknowledges the usage of GeoGebra \cite{geogebra} as an interactive tool, which was used to explore and visualise iterative variants of the response curves during the development of the compact model, shown in \autoref{eq:lvdt-eq}.

Special thanks are extended to the departmental engineer,  Wim Beaumont, for his technical assistance and support during the experimental. The author also sincerely acknowledges Dr. Hans Van Haevermaet, for guidance, supervision, and scientific discussions throughout this work.

The author further acknowledges Dr. Anoop Nagesh Koushik for developing and maintaining the software framework used for acquiring measurement data from the DAQ systems employed in this study.

\section*{Data availability}
The finite-element simulation pipeline developed for this work is available through the author’s Git repository: \url{https://github.com/beginner117/LVDT-VC-modelling-toolkit/releases/tag/v1.0 } (accessed on 19 January 2026), which corresponds to the version used in this paper.
The experimental datasets and additional processed data supporting the findings of this study are available from the corresponding author upon request for further analysis and verification.

\bibliographystyle{unsrt}
\bibliography{references}
% \include{references.bib}
% \printbibliography

\end{document}